\documentclass[runningheads]{llncs}
\usepackage{array}
\usepackage{multirow}
\usepackage{graphicx}
\usepackage{booktabs}
\usepackage{amsmath}
\usepackage{url}
\usepackage[utf8]{inputenc}
\usepackage{enumitem}
\usepackage{tabularx}
\usepackage{ragged2e}
\usepackage[T1]{fontenc}
\usepackage{tabularx}
\usepackage{amsmath}
\usepackage{listings}
\usepackage{xcolor}
\usepackage{hyperref}
\usepackage{algorithm}
\usepackage{algpseudocode}
\usepackage{xspace}

\usepackage{graphicx}
\definecolor{tlacomment}{rgb}{0.0,0.6,0.0}
\definecolor{tlakeyword}{rgb}{0.0,0.0,1.0}
\definecolor{tlastring}{rgb}{0.58,0.0,0.82}

\lstdefinelanguage{tla}{
  morekeywords={MODULE, EXTENDS, CONSTANTS, VARIABLES, ASSUME, LET, IN, IF, THEN, ELSE, CASE, OTHER, WITH, EXCEPT, RECURSIVE, LOCAL, INSTANCE, THEOREM, LEMMA, COROLLARY, PROPOSITION, PROOF, OBVIOUS, QED, BY, DEF, USE, HIDE, PICK, SUFFICES, UNION, DOMAIN, SUBSET, ENABLED, UNCHANGED, \A, \E, \X, \cdot, ~>, ->, <-, <<, >>, ==, /=, <=, >=, .., +-, -+},
  sensitive=true,
  morecomment=[l]{(*},
  morecomment=[n]{*)},
  morestring=[b]",
  keywordstyle=\color{tlakeyword},
  commentstyle=\color{tlacomment},
  stringstyle=\color{tlastring}
}
\lstdefinelanguage{cypher}{
  morekeywords={
    MATCH, RETURN, WHERE, CREATE, DELETE, SET, DETACH, MERGE, ON, OPTIONAL,
    AS, AND, OR, NOT, IN, STARTS, ENDS, CONTAINS, DISTINCT, ORDER, BY, LIMIT, SKIP
  },
  sensitive=true,
  morecomment=[l]{//},
  morestring=[b]",
  keywordstyle=\color{blue}\bfseries,
  commentstyle=\color{gray},
  stringstyle=\color{red},
}
\begin{document}
\title{Alloy-Driven Verification of Object-Centric Event Data: From Temporal Logic to Knowledge Graphs}
\titlerunning{From Temporal Logic to Knowledge Graphs}
\author{}
\author{
Saba Latif\inst{1} \and
Huma Latif\inst{2} \and
Touseef Ur Rehman\inst{3} \and
Muhammad Rameez Ur Rahman\inst{4}
}

\authorrunning{S. Latif et al.}

\institute{
 Sapienza University of Rome, Italy\inst{1} \quad
 University of Sahiwal, Pakistan\inst{2} \quad
 East China University of Science and Technology, China\inst{3} \quad
 Ca' Foscari University of Venice, Italy\inst{4}\\[3pt]
\textit{Emails:} \texttt{latif@di.uniroma1.it, latifhuma666@gmail.com, touseef@mail.ecust.edu.cn, muhammad.rahman@unive.it}
}

\maketitle              
\begin{abstract}
Object-centric process mining addresses the limitations of traditional approaches, which often involve the lossy flattening of event data and obscure vital relationships among interacting objects. This paper presents a novel formal framework for Object-centric Event Data (OCED) that ensures the correctness of the meta-model and preserves native object-centric semantics prior to the system implementation. Our approach effectively leverages Alloy for precisely specifying temporal properties and structural relationships between objects and events. This guarantees thorough verification against predefined OCED constraints such as cross-object cardinality bounds and time-aware consistency rules, hence preventing common data integrity issues. We demonstrate the effectiveness of the proposed framework in discovering and validating implicit object dependencies in event logs, particularly when importing data into graph databases like Neo4j. This demonstrates how formal verification can avoid pitfalls that lead to data invisibility and improve knowledge graph creation, enrichment, and querying.
To bridge theory and practice, our verified \emph{FOCED} is made accessible through automatically generated Python bindings, empowering industrial users without formal methods expertise. The code is available on GitHub \footnote{\url{https://github.com/sabalati/FOCED}}

\end{abstract}

\section{Introduction}
Process mining techniques perform crucial tasks, including discovery, monitoring, improvement, and analysis of event data to extract insights from event logs. \cite{van2016data}. Traditional techniques oversimplify processes by restricting each event to a single case, ignoring interactions among multiple related objects. Object-centric process mining resolves this by enabling the analysis of interdependent processes spanning multiple business units\cite{vanDerAalst2019}. The Object-Centric Event Log (OCEL) 1.0 format \cite{OCEL2020} was introduced as a standard for storing and exchanging object-centric event data; however, its limited expressiveness, such as the lack of object change tracking, has led to calls for enhancement. We use OCEL logs as input to our model to broaden the scope of our work.
In graph databases like Neo4j\footnote{\url{https://neo4j.com/}}, event-object logs often contain implicit relationships.
The limitations of current standards, such as IEEE 1849-2016 eXtensible Event Stream(XES)\cite{xes2016ieee} have become increasingly apparent as organizations attempt to analyze complex, object-rich processes. In response, the process mining community has developed object-centric approaches, culminating in the ongoing development of the OCED meta-model \cite{oced2023}. We adopt the same OCED model but extend it through the application of formal methods.
\\ 
Formal methods provide a mathematically rigorous framework for specifying, designing, and verifying systems, enabling early error detection to reduce costly failures and vulnerabilities \cite{latif2019smart,latif2017survey}. Key techniques include model checking and theorem proving \cite{latif2018intelligent}. In this work, the authors adopt model checking for its automation and scalability, particularly in verifying blockchain consensus protocols \cite{bertrand2024reusable}. 
In this paper, we propose Formal Object-Centric Event Data (\emph{FOCED}), an extension of flattened event logs that adopts an object-centric approach. 
Our work is inspired from Latif et al. \cite{Latif2025ASemantic}, introducing a semantic reference implementation of the OCED meta-model. Their framework enhances OCED using Semantic Web technologies, adding an \emph{Intensional Level} for embedding domain knowledge into the core meta-model and an \emph{Extensional Level} that instantiates these abstractions as knowledge graphs populated with process data \cite{Latif2025ASemantic}. 

Our work introduces a formalized method for transforming flat, structured data into accurate Formal Object-Centric Event Data (\emph{FOCED}), enabling robust multi-object process analysis beyond existing approaches. Building on \cite{Latif2025ASemantic} and \cite{Latif2025ObjectCentric}, it advances object-centric process mining through three core contributions: a detailed analysis of limitations in case-oriented models with interoperability across XES and OCEL, the establishment of formal foundations for OCED to represent complex processes, and a reference Alloy-based implementation that demonstrates practical applicability and supports reproducibility and industrial adoption.

The rest of the paper is organized as follows: Section 2 reviews related work, Section 3 outlines the Problem Statement and Framework Overview, Section 4 presents the Cypher Query Analysis of \emph{FOCED} Knowledge Graph in Neo4j, and Section 5 concludes with future directions.

\section{Related Work}
In this section, we review existing literature on the technical challenges of object-centric process mining, the development of formal specifications for event data, and efforts to build practical tools and frameworks in this domain.
OCED links events to multiple business objects, capturing richer context than traditional case-based logs. Since 2020, key models include the OCEL~\cite{ghahfarokhi2021ocel}, OCEDO and the Event Knowledge Graph (EKG)~\cite{jalali2020}, alongside formats like XOC~\cite{li2018extracting}. While these enable techniques such as precision measurement~\cite{adams2021precision}, variant analysis~\cite{adams2022defining}, and anomaly filtering~\cite{berti2022filtering}, challenges remain in modeling evolving attributes, ensuring clear object-attribute relationships, and protecting sensitive data. OCEL~2.0~\cite{ocel2} advances inter-object relationship modeling and dynamic attribute tracking, but constructing accurate models from real-world, incomplete, and ambiguous logs~\cite{tax2022,van2022multi} remains difficult. Existing tools such as OC-PM~\cite{berti2023ocpm}, ocpa~\cite{adams2022ocpa}, and PM4Py object-centric extensions~\cite{berti2023pm4py} support analysis but still rely on well-structured inputs. 
The proposed model, validated the BPIC 2013 dataset, enables highly flexible and complex analyses via SPARQL queries, illustrated by its ability to precisely identify nuanced process behaviors, such as inter-team transfer "ping-pong" based on OCEDO Knowledge graph \cite{Latif2025ObjectCentric}. But it needs to be formally verified to check its correctness. 

Formal methods have been successfully applied in domains such as IoT, smart cities, and critical infrastructure, utilizing tools like UML, automata theory, and TLA+ for verification \cite{latif2018intelligent}. In process mining, LTL-based preprocessing has been used to enhance event log quality and capture dynamic states, such as fluctuating product prices in e-commerce, which are essential for accurate trend analysis \cite{vanDerAalst2005ProcessMining}. 
Recent advances combine OCED with formal methods for precise, verifiable multi-object process analysis \cite{kretzschmann2025data}. Our work builds on these foundations, providing a reference implementation that bridges OCED with formal verification to improve accuracy, reliability, and analytical depth in object-centric process mining. 
\section{Problem Statement and Framework Overview} 
Our approach addresses two key challenges in process mining: data integrity and analytical depth. We formally specify the \emph{FOCED} model in Alloy \cite{jackson2002alloy} to verify structural and temporal soundness, ensuring a consistent and validated event foundation. The verified model is then transformed into a Knowledge Graph \cite{rodriguez2020knowledge}, enabling Cypher-based analysis of complex, multi-object relationships that conventional case logs cannot capture. This integration of formal verification and graph analytics provides a reliable and expressive basis for object-centric process mining, aligned with the OCED specification \cite{oced2023}.
\begin{figure}[t]
    \includegraphics[width=1\linewidth]{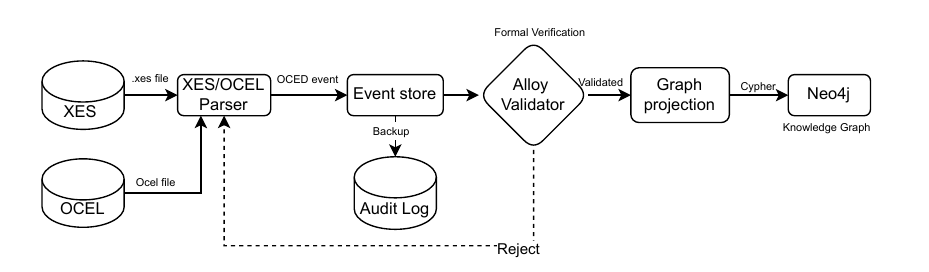}
        \caption{Process flow of Formal OCED (\emph{FOCED})}
    \label{fig:PF}
\end{figure}
In Figure \ref{fig:PF}, we illustrate a workflow involving event log processing, formal verification, and knowledge graph construction. The components are organized as follows:

\begin{itemize}
    \item \textbf{XES/OCEL Log}: Represents an event log file in the XES or OCEL format, commonly used for storing process-related event data.
    \item \textbf{XES/OCEL Parser}: A Python pm4py module responsible for parsing and interpreting files to extract event data.
    \item \textbf{OCED Event}: Refers to an event within an OCED model, which extends traditional event logs to handle multiple object interactions.
    \item \textbf{Event Store}: A database or storage system where events are persisted for further analysis or retrieval.
    \item \textbf{Backup}: Indicates a backup mechanism for the event data or system state.
    \item \textbf{Audit Log}: A log for recording system activities or changes, often for compliance or debugging purposes.
    \item \textbf{Formal Verification}: The process of rigorously verifying system correctness using mathematical methods.
    \item \textbf{Alloy Validator}: A tool for validating system models specified in Alloy, a formal specification language.
    \item \textbf{Validated}: Confirms that the system or data has undergone successful verification.
    \item \textbf{Graph Projection}: The transformation of event data into a graph structure for analysis.
    \item \textbf{Cypher}: A query language used for interacting with graph databases, such as Neo4j.
    \item \textbf{Neo4j}: A popular graph database management system.
    \item \textbf{Knowledge Graph}: A structured representation of knowledge, often derived from event data, enabling advanced querying and insights.
\end{itemize}

In this pipeline, event logs are parsed, stored, and formally verified before being projected into a knowledge graph for analysis. This workflow is novel, integrating process mining, formal methods, and knowledge-driven applications. PlantUML state diagram can be accessed using following link.\footnote{Generated using PlantUML: \url{https://tinyurl.com/pplantuml}}
\begin{figure}[t]
    \includegraphics[width=0.9\linewidth]{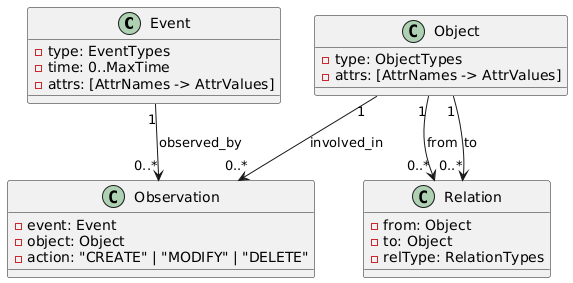}
    \caption{\emph{FOCED} Model Visualization using  PlantUML}
    \label{fig:FM}
\end{figure}
\begin{table}[t]
\centering
\scriptsize
\setlength{\tabcolsep}{5pt}
\renewcommand{\arraystretch}{1.1}
\begin{tabularx}{\linewidth}{@{}lX@{}}
\toprule
\textbf{Aspect} & \textbf{Description} \\ 
\midrule
\textbf{Name / Purpose} & OCED Meta-Model Initialization \& Event Handling for constructing and evolving stateful systems via the event–object paradigm. \\
\textbf{Core Components} & \texttt{events}, typed \texttt{objects}, \texttt{relations}, and \texttt{observes} (MODIFY links). \\
\textbf{Inputs / Outputs} & Inputs: \texttt{EventTypes}, \texttt{ObjectTypes}, \texttt{AttrNames/Values}, \texttt{MaxTime}, \texttt{MaxObserves}, \texttt{RelationTypes}. 
Outputs: consistent \{\texttt{events}, \texttt{objects}, \texttt{relations}, \texttt{observes}\}. \\
\textbf{Operations} & \texttt{AddObject(type, attrs)}, \texttt{AddEvent(type, time, attrs, linkedObjs)}. \\
\textbf{Constraints} & Valid types/times; $|\texttt{linkedObjs}| \leq \texttt{MaxObserves}$; valid attributes. \\
\textbf{Use Cases} & Simulation, compliance testing, prototyping, and event log generation. \\
\bottomrule
\end{tabularx}
\caption{Overview of the \emph{FOCED}, defining object, event, and observation structures together with incident lifecycle and cardinality constraints.}
\label{tab:oced-summary}
\vspace{-2mm}
\end{table}

\subsection{Formal Verification Using Alloy}
\begin{figure}[t]
    \centering
    \includegraphics[width=0.8\linewidth, height=5.5cm]{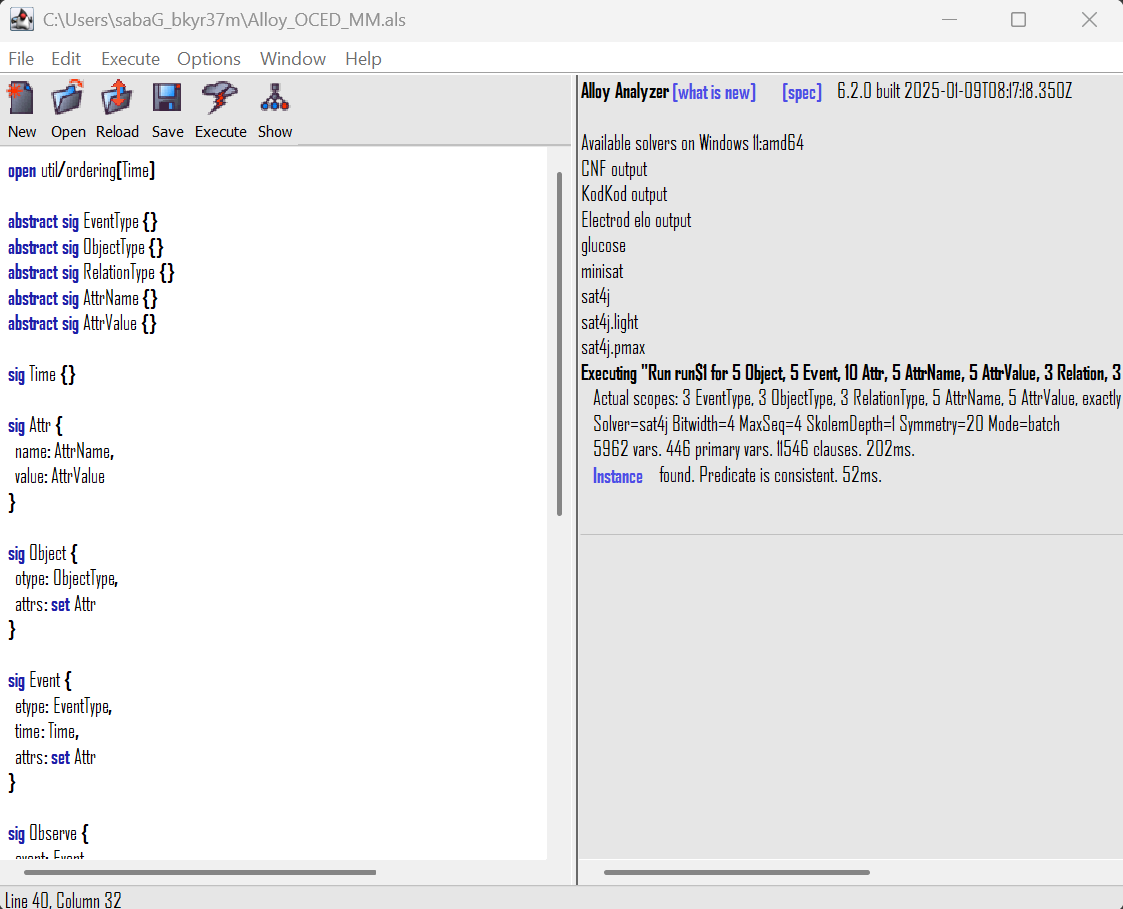}
    \caption{Verified Alloy \emph{FOCED} satisfying all constraints}
    \label{fig:Con}
\end{figure}

\begin{figure}[t]
    \centering
    \includegraphics[width=1.0\linewidth, height=7.0cm]{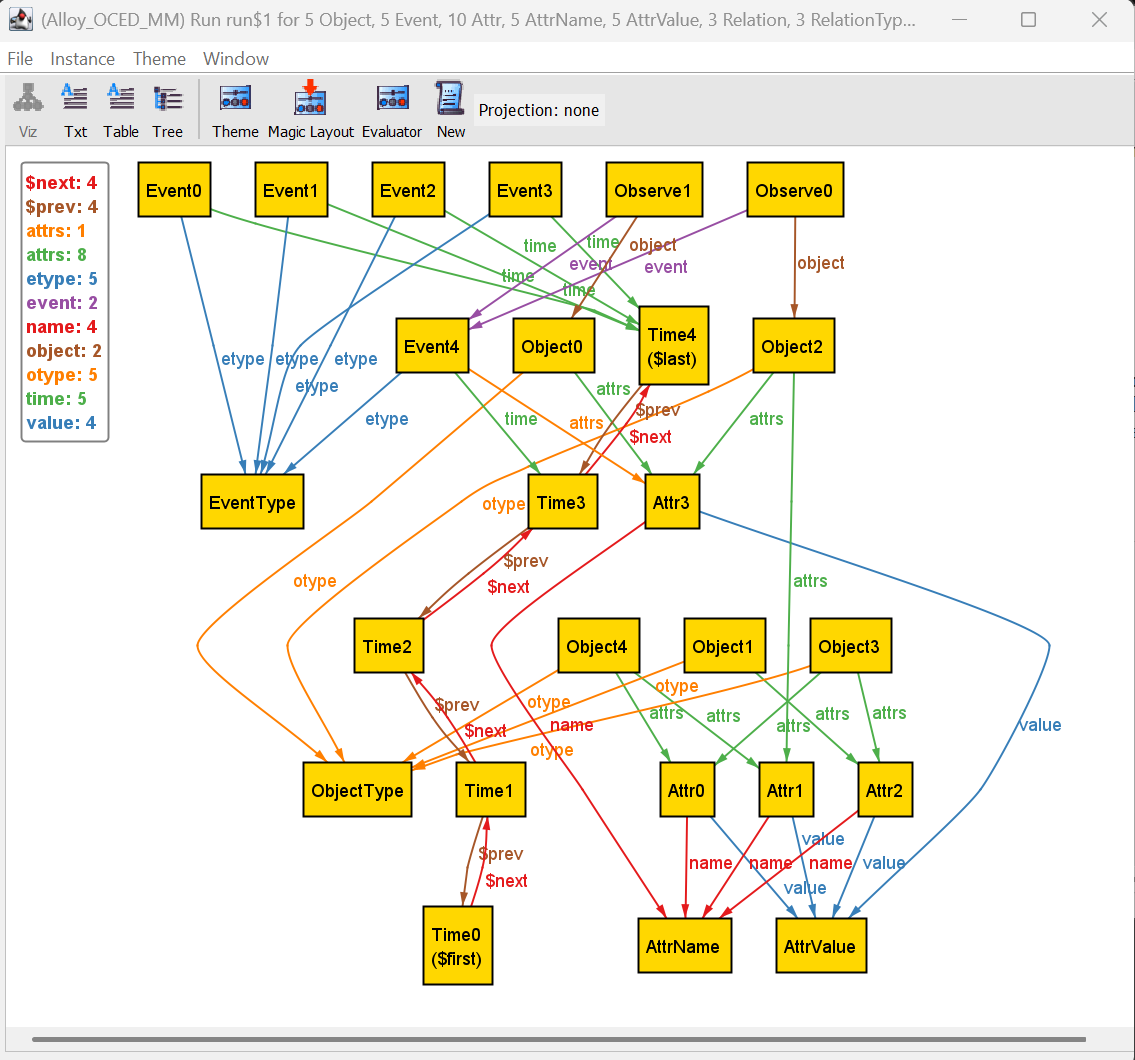}
    \caption{Instances generated by Alloy satisfying all constraints of the proposed \emph{FOCED}.}
    \label{fig:in}
\end{figure}

To ensure the correctness of our \emph{FOCED} model, we used the Alloy Analyzer to formally verify key properties. The core OCED model is encoded in Alloy, specifying events, objects, relations, and observations, along with constraints such as attribute validity and a bound on the number of objects observed per event\footnote{\url{https://github.com/sabalati/FOCED/blob/main/FOCED_Alloy_MM.als}}. \emph{FOCED} with enhanced Alloy constraints for incident-centric event–object systems\footnote{\url{https://github.com/sabalati/FOCED/blob/main/Enhanced_FOCED_Constraints.als}}. One of the critical properties verified was:
\begin{quote}
\emph{For every event, the number of observed objects does not exceed a predefined maximum (MaxObserves).}
\end{quote}
This was encoded as an assertion in Alloy:
\begin{verbatim}
assert MaxObserveProperty {
  all e: Event | 
    let objs = { o: Object | some obs: Observe \end{verbatim} \begin{verbatim}| obs.event = e and obs.object = o } |
      #objs <= MaxObserves
}check MaxObserveProperty for 5\end{verbatim}
The Alloy Analyzer confirmed that this property holds by reporting, and it confirms that our model is verified on the given properties \ref{fig:Con}:
\begin{quote}
\emph{No counterexample found. Assertion may be valid.}
\end{quote}
This confirms that our model respects the specified bound across all examined scopes, providing confidence in the integrity of our design as shown in Figure \ref{fig:in}.
\section{Cypher Query Analysis of \emph{FOCED} Knowledge Graph in Neo4j}
The transformation of \emph{FOCED} validated data into a Knowledge Graph enables powerful, relationship-based Cypher queries. This enables the precise investigation of complex, multi-object interactions and causal chains, tasks that are impossible with traditional methods. By providing a verified model and a flexible graph structure, our work allows for a deeper and more accurate understanding of process dynamics. Initial results using BPIC 2013 dataset as XES input are shown as Neo4j graphs with Cypher query outputs. The BPIC 2013 dataset, derived from Volvo IT Belgium’s VINST incident and problem-management system, contains over 9,000 cases and 74,000 events capturing status, assignment, and escalation activities across multiple support teams \footnote{\url{https://data.4tu.nl/articles/dataset/BPI_Challenge_2013_incidents/12693914}}.
\begin{figure}[t]
    \centering
    \includegraphics[width=1.0\linewidth, height=5.8cm]{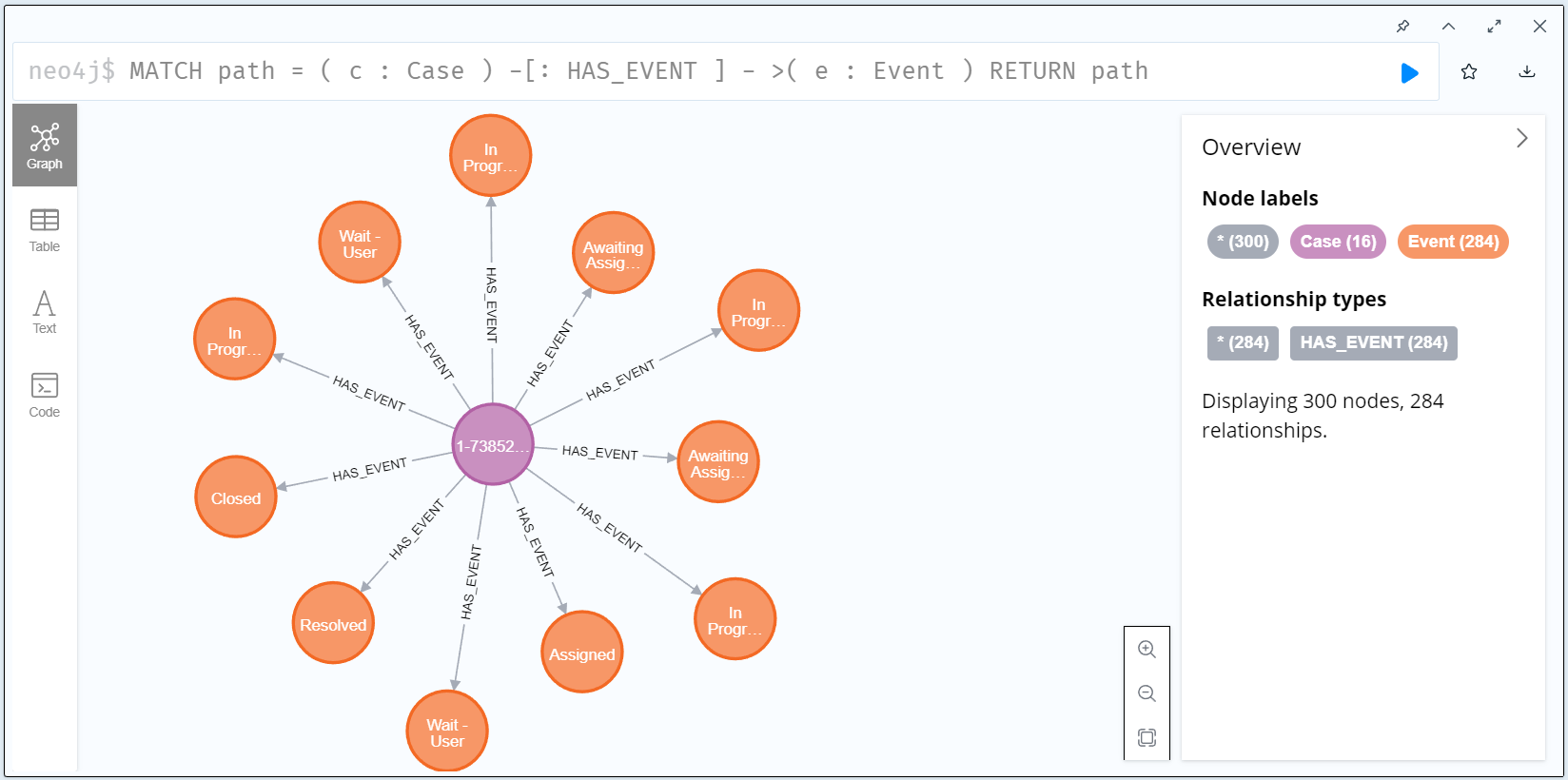}
    \caption{Example visualization of case-event paths in Neo4j showing case nodes connected to event nodes}
    \label{fig:caseevent}
\end{figure}
\begin{figure}[!htb]
    \centering
\includegraphics[width=1.0\linewidth]{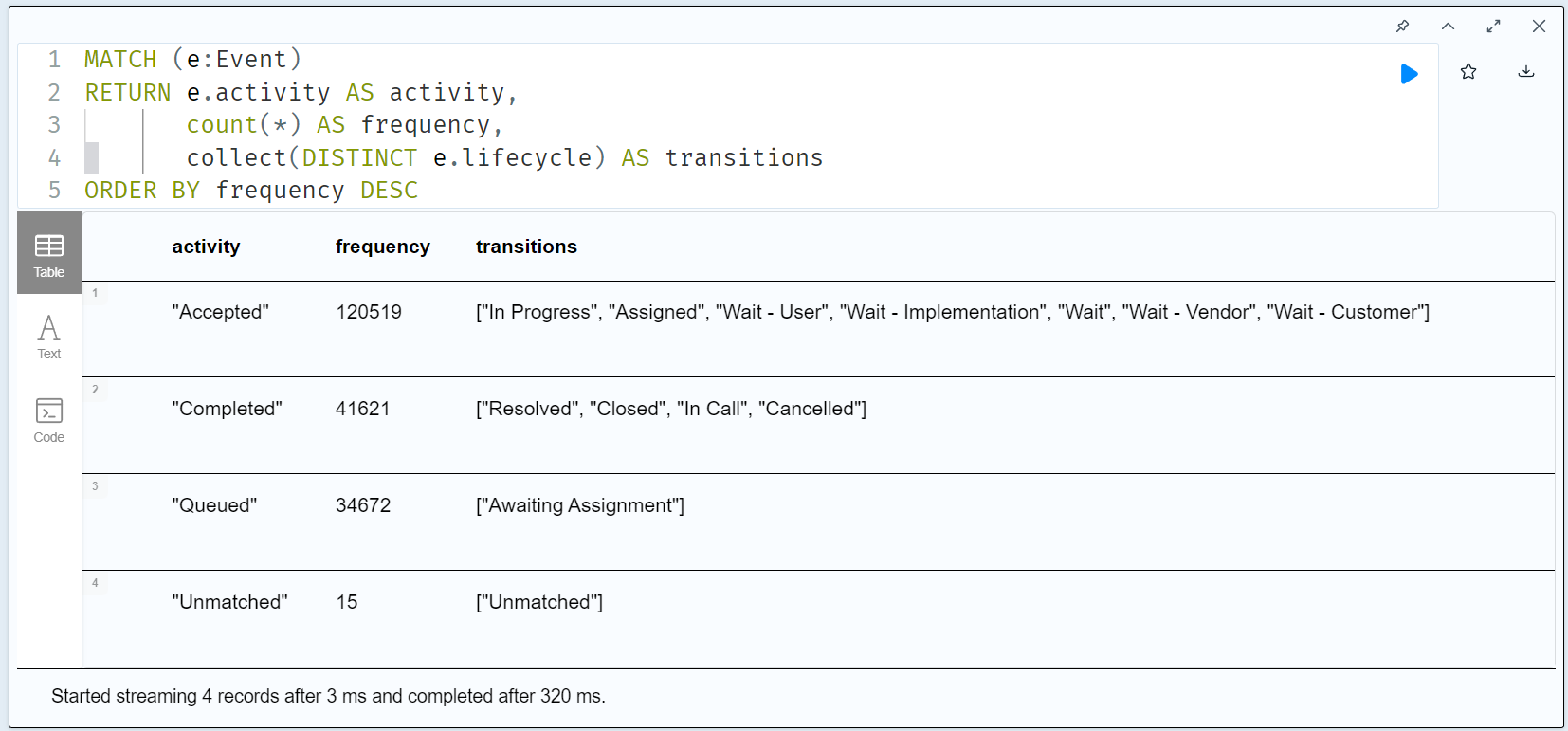}
    \caption{Visualization of activity frequencies with their distinct lifecycle transitions, ranked by occurrence count}
    \label{fig:query2}
\end{figure}
\subsection{Graph Path Visualization Query}
\begin{lstlisting}[language=SQL]
MATCH path = (c:Case)-[:HAS_EVENT]->(e:Event) RETURN path
\end{lstlisting}
This fundamental query retrieves and visualizes the complete event log structure stored in Neo4j by:
\textbf{Pattern Matching}: Identifies all paths connecting:
Case nodes (\texttt{(c:Case)}) representing process instances, Event nodes (\texttt{(e:Event)}) containing activity executions, Connected via \texttt{HAS\_EVENT} relationships, \textbf{Path Composition}: Each returned path represents: One complete case-to-event connection and Visualized as directed arrows in Neo4j as shown in Figure\ref{fig:caseevent}.
The analytical value of the path query lies in its ability to verify multiple aspects of the event log. It ensures that the XES import structure is correctly formed (\emph{Data Structure}), reveals overall case flow patterns (\emph{Process Discovery}), detects any disconnected components or anomalies in the log (\emph{Data Quality}), and validates that the recorded events follow a correct temporal sequence (\emph{Temporal Order}).

\subsection{Activity Frequency Analysis Query}
This query analyzes activity frequency in our event log by counting occurrences of each activity, collecting all distinct lifecycle transitions per activity, and sorting results by frequency (most common first).
It identifies the most/least frequent activities, reveals activities with multiple transition types (e.g., start/complete), helps spot potential bottlenecks through high-frequency activities, and validates proper event capture in XES imports as shown in Figure \ref{fig:query2}.
\begin{lstlisting}[language=SQL]
MATCH (e:Event)
RETURN e.activity AS activity,
       count(*) AS frequency,
       collect(DISTINCT e.lifecycle) AS transitions
ORDER BY frequency DESC
\end{lstlisting}

\subsection{Basic Event Sequence Query}
\begin{lstlisting}[language=Cypher]
MATCH (c:Case)-[:HAS_EVENT]->(e:Event)
RETURN c.id AS case_id, 
       e.activity AS activity, 
       e.lifecycle AS transition, 
       e.timestamp AS timestamp
ORDER BY c.id, e.timestamp
\end{lstlisting}
\begin{figure}[t]
    \centering
    \includegraphics[width=0.8\linewidth, height=6cm]{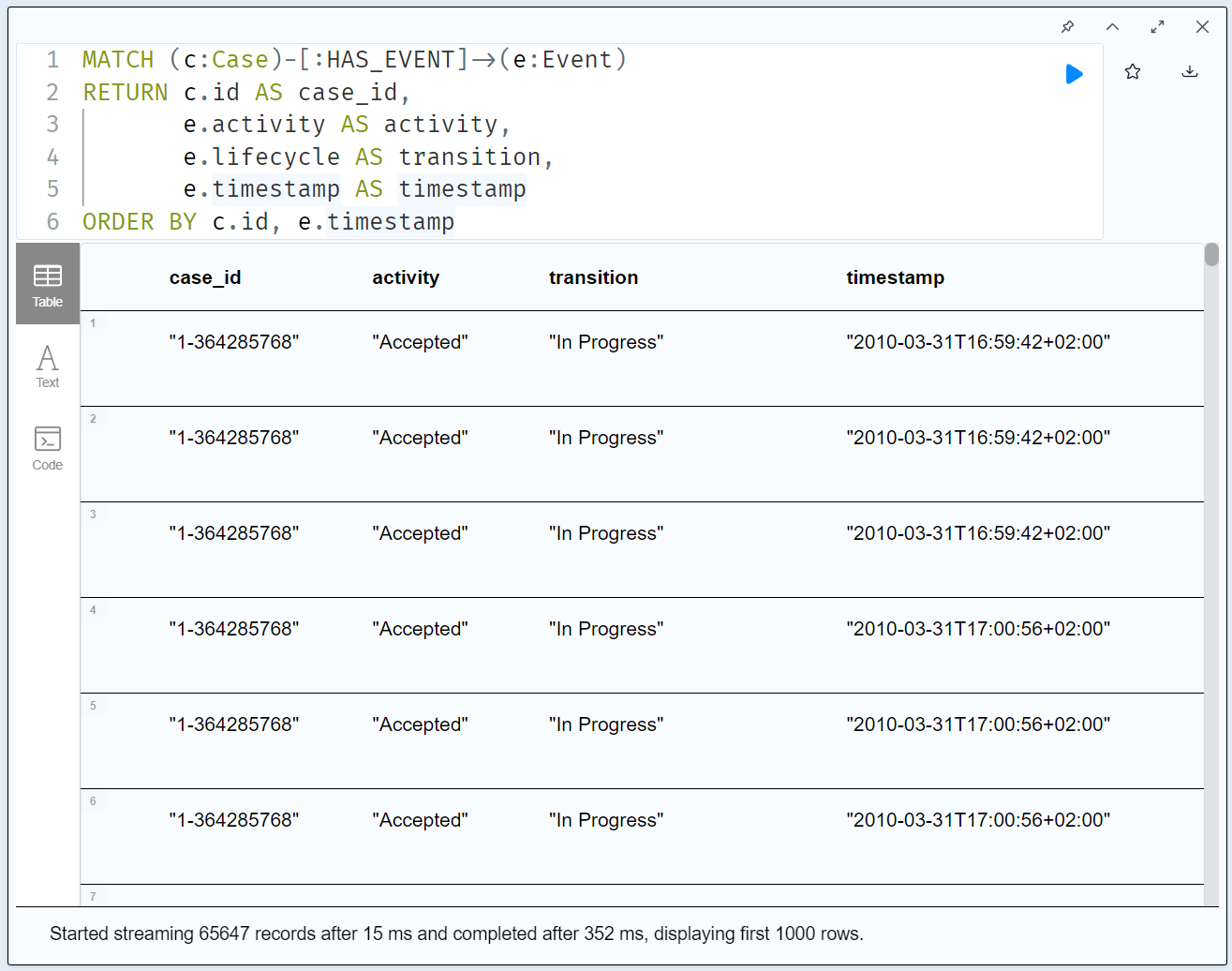}
    \caption{Visualization of activity frequencies showing distribution across the process}
    \label{fig:A}
\end{figure}
The foundational query retrieves all cases along with their associated events, returning activities, lifecycle transitions, and timestamps in temporal order per case. This verifies the integrity of XES imports, event sequencing, and case event consistency, forming the basis for subsequent analyses, as shown in Figure~\ref{fig:A}. Evaluation metrics encompass data completeness, process discovery, variability, temporal correctness, and import quality, confirming the accuracy of the XES-to-Neo4j conversion. Using Neo4j\footnote{Neo4j Graph Database: \url{https://neo4j.com/}} and its APOC plugin\footnote{APOC Library: \url{https://neo4j.com/labs/apoc/4.1/installation/}}, the \emph{FOCED} formalization is validated through graph-based reasoning, ensuring cross-object and time-aware consistency. The Python pipeline provides automation and accessibility. Key contributions include dynamic attribute handling, explicit object relationship modeling, and scalable event log processing, demonstrated in Figures~\ref{fig:Con} and~\ref{fig:in}. The same framework supports OCEL compliance, though only XES-based results are shown here.

\subsection{Framework's Contribution and Validation Comparison with other Reference Implementations}
Our framework enhances process mining by integrating formal methods to ensure data integrity. Event logs are represented as labeled property graphs, with business rules encoded as Linear Temporal Logic (LTL) properties. A formal verification step checks whether the event log satisfies all LTL rules, and any violations are precisely localized, enabling direct identification of data integrity issues. To illustrate this, we compare our approach with OCEL \cite{ghahfarokhi2021ocel} and the Event Knowledge Graph (EKG) \cite{ekg2020}, focusing on data models, query capabilities, and extensibility. Table~\ref{tab:comparison} summarizes these similarities and differences, highlighting the strengths of each approach for event-centric applications.

\vspace{-6pt}
\subsection{Temporal Logic Specifications}
\vspace{-4pt}

We formalize incident lifecycle constraints from the BPI Challenge 2013 log as Linear Temporal Logic (LTL)-inspired specifications derived from the \emph{FOCED} model and aligned with ITSM semantics. These are evaluated through Cypher-based temporal pattern queries within Neo4j, serving as executable approximations of formal LTL rules.
\textbf{Safety:} Timestamps and assignments remain valid for known incidents.
\textbf{Cardinality:} Escalations are limited to $\leq3$ steps before resolution.
\textbf{Liveness:} Every incident eventually reaches a closed state, and reopened cases are again resolved.
\textbf{Fairness:} Operator updates precede valid status transitions.
\textbf{Consistency:} Priority coherently reflects impact and urgency.
These constraints establish a verifiable baseline for automated compliance, lifecycle integrity, and performance validation within the Neo4j graph.
\begin{table}[t]
\setlength{\tabcolsep}{3pt}
\renewcommand{\arraystretch}{0.85}
\centering
\caption{Compact comparison of OCEL, EKG, Semantic OCED, and \emph{FOCED}}
\label{tab:comparison}
\scriptsize
\begin{tabularx}{\linewidth}{|l|X|X|X|X|}
\hline
\textbf{Feature} & \textbf{OCEL \cite{ghahfarokhi2021ocel}} & \textbf{EKG \cite{ekg2020}} & \textbf{Semantic OCED \cite{Latif2025ASemantic}} & \textbf{\emph{FOCED}} \\
\hline
Model & Object-centric log & RDF/OWL KG & OCEDO OCEDD
OCEDR & Labeled KG (Neo4j) \\
\hline
Storage & JSON/XML & RDF store & RDF/OWL (Turtle, RDF/XML) & Neo4j DB \\
\hline
Input & OCEL JSON/XML & RDF vocab. & Process data & XES/OCEL via pm4py/XML \\
\hline
Event–Case Link & Object refs. & Ontological rels. & Semantic rels. (\texttt{oced:observes}) & (Event)-[:INVOLVES]->(Object) \\
\hline
Time & Timestamp attr. & Temporal props. & \texttt{oced:observed\_at} (\texttt{xsd:dateTime}) & ISO timestamp on Event \\
\hline
Query & Python (pm4py) & SPARQL & SPARQL/RDF tools & Cypher + APOC \\
\hline
Tooling & OCEL-Tools & RDF libs & RDF/OWL editors, RML & Neo4j, APOC, Python \\
\hline
Semantics & OCEL schema & Ontologies & 3-layer ontology & Graph + Alloy-style rules \\
\hline
Customize. & Medium & High & Very high (extensible RDF) & High (schema-on-read) \\
\hline
Strengths & Standard, simple & Semantic power & Semantics interoperability & Speed, flexibility, constraints \\
\hline
Weaknesses & Low expressiveness & Complex, heavy & - & Non-standard semantics \\
\hline
\end{tabularx}
\end{table}

\section{Conclusion}
This work introduced a formal verification framework for object-centric process mining, ensuring the structural and temporal correctness of \emph{FOCED} models. Building on Alloy’s declarative precision, the framework formalizes \emph{FOCED} to guarantee compliance with meta-model constraints such as cross-object cardinality and time-aware consistency before implementation. Coupling formal verification with process mining enables the discovery and validation of implicit object dependencies and prevents integrity issues that often arise during data import into graph databases like Neo4j. The verified model strengthens knowledge graph creation, enrichment, and querying while providing a rigorous bridge between theory and practical application.

While current validation is based on XES input, the framework also supports OCEL. Future work will focus on cross-format event log analysis, incremental discovery algorithms, enhanced visualization and analytics, as well as domain-specific extensions. Moreover, we aim to integrate Large Language Models (LLMs) driven agents to automate rule translation, Cypher query generation, and OCED data mapping, improving accessibility and usability for non-expert users across industrial and research settings.
\section{Acknowledgement}
The work of Saba Latif received funding from MUR under PRIN programme, grant B87G22000450001 (PINPOINT).

%
%

\bibliographystyle{splncs04}
\bibliography{bib}
\end{document}